\documentclass[twocolumn,tighten,times]{aastex631}

\usepackage[T1]{fontenc}

\newcommand\ergs{erg~s$^{-1}$}

\shorttitle{X-ray Polarimetry of Sco X-1 with PolarLight}
\shortauthors{Long et al.}

\graphicspath{{./}{fig/}}
 
\begin{document}
 
\title{A significant detection of X-ray Polarization in Sco X-1 with PolarLight and constraints on the corona geometry}

\author{Xiangyun Long}
\affiliation{Department of Engineering Physics, Tsinghua University, Beijing 100084, China}

\correspondingauthor{Hua Feng}
\email{hfeng@tsinghua.edu.cn}

\author[0000-0001-7584-6236]{Hua Feng}
\affiliation{Department of Astronomy, Tsinghua University, Beijing 100084, China}
\affiliation{Department of Engineering Physics, Tsinghua University, Beijing 100084, China}

\author{Hong Li}
\affiliation{Department of Astronomy, Tsinghua University, Beijing 100084, China}

\author{Jiahuan Zhu}
\affiliation{Department of Astronomy, Tsinghua University, Beijing 100084, China}

\author{Qiong Wu}
\affiliation{Department of Engineering Physics, Tsinghua University, Beijing 100084, China}

\author{Jiahui Huang}
\affiliation{Department of Engineering Physics, Tsinghua University, Beijing 100084, China}

\author{Massimo Minuti}
\affiliation{INFN-Pisa, Largo B. Pontecorvo 3, 56127 Pisa, Italy}

\author{Weichun Jiang}
\affiliation{Key Laboratory for Particle Astrophysics, Institute of High Energy Physics, Chinese Academy of Sciences, Beijing 100049, China}

\author{Dongxin Yang}
\affiliation{Department of Engineering Physics, Tsinghua University, Beijing 100084, China}

\author{Saverio Citraro}
\affiliation{INFN-Pisa, Largo B. Pontecorvo 3, 56127 Pisa, Italy}

\author{Hikmat Nasimi}
\affiliation{INFN-Pisa, Largo B. Pontecorvo 3, 56127 Pisa, Italy}

\author{Jiandong Yu}
\affiliation{School of Electronic and Information Engineering,  Ningbo University of Technology, Ningbo, Zhejiang 315211, China}

\author{Ge Jin}
\affiliation{North Night Vision Technology Co., Ltd., Nanjing 211106, China}

\author{Ming Zeng}
\affiliation{Department of Engineering Physics, Tsinghua University, Beijing 100084, China}

\author{Peng An}
\affiliation{School of Electronic and Information Engineering,  Ningbo University of Technology, Ningbo, Zhejiang 315211, China}

\author{Jiachen Jiang}
\affiliation{Department of Astronomy, Tsinghua University, Beijing 100084, China}

\author{Enrico Costa}
\affiliation{IAPS/INAF, Via Fosso del Cavaliere 100, 00133 Rome, Italy}

\author{Luca Baldini}
\affiliation{INFN-Pisa, Largo B. Pontecorvo 3, 56127 Pisa, Italy}

\author{Ronaldo Bellazzini}
\affiliation{INFN-Pisa, Largo B. Pontecorvo 3, 56127 Pisa, Italy}

\author{Alessandro Brez}
\affiliation{INFN-Pisa, Largo B. Pontecorvo 3, 56127 Pisa, Italy}

\author{Luca Latronico}
\affiliation{INFN, Sezione di Torino, Via Pietro Giuria 1, I-10125 Torino, Italy}

\author{Carmelo Sgr\`{o}}
\affiliation{INFN-Pisa, Largo B. Pontecorvo 3, 56127 Pisa, Italy}

\author{Gloria Spandre}
\affiliation{INFN-Pisa, Largo B. Pontecorvo 3, 56127 Pisa, Italy}

\author{Michele Pinchera}
\affiliation{INFN-Pisa, Largo B. Pontecorvo 3, 56127 Pisa, Italy}

\author{Fabio Muleri}
\affiliation{IAPS/INAF, Via Fosso del Cavaliere 100, 00133 Rome, Italy}

\author{Paolo Soffitta}
\affiliation{IAPS/INAF, Via Fosso del Cavaliere 100, 00133 Rome, Italy}


\begin{abstract}

We report the detection of X-ray polarization in the neutron star low mass X-ray binary Scorpius (Sco) X-1 with PolarLight. The result is energy dependent, with a non-detection in 3--4 keV but a 4$\sigma$ detection in 4--8 keV; it is also flux dependent in the 4--8 keV band, with a non-detection when the source displays low fluxes but a 5$\sigma$ detection during high fluxes, in which case we obtain a polarization fraction of $0.043 \pm 0.008$ and a polarization angle of $52\fdg6 \pm 5\fdg4$. This confirms a previous marginal detection with OSO-8 in the 1970s, and marks Sco X-1 the second astrophysical source with a significant polarization measurement in the keV band.  The measured polarization angle is in line with the jet orientation of the source on the sky plane~(54\arcdeg), which is supposedly the symmetric axis of the system.  Combining previous spectral analysis, our measurements suggest that an optically thin corona is located in the transition layer under the highest accretion rates, and disfavor the extended accretion disk corona model. 

\end{abstract} 

\section{Introduction}

Sco X-1 is the first discovered extrasolar X-ray source \citep{Giacconi1962} and the brightest persistent object in the keV sky besides the Sun. It is a low mass X-ray binary (LMXB) containing a neutron star \citep{Steeghs2002} with an orbital period of 0.787~d \citep{Galloway2014} at a distance of 2.13~kpc \citep{Arnason2021}. The source is classified to be a so-called Z source based on its evolutionary pattern on the color-color diagram \citep{Hasinger1989}.  Sco X-1 represents a class of LMXBs with high accretion rates, with a peak luminosity of around $2 \times 10^{38}$~\ergs, which is close to the Eddington limit of a 1.4~$M_\odot$ neutron star \citep{Titarchuk2014}.  Unlike black holes, the hard surface of neutron stars will stop the accretion flow and dissipate the energy that can otherwise be swallowed by the event horizon. Therefore, a transition layer, or similarly a spreading layer or a boundary layer, is expected in between the Keplerian accretion disk and the star surface \citep{Inogamov1999}. The existence of such a layer is supported by X-ray timing \citep{Sunyaev2000}. 

X-ray studies of Sco X-1 have been conducted mainly with the Rossi X-ray Timing Explorer (RXTE) in the energy range from about 3~keV to a few hundred keV, because the extreme brightness of the source may saturate most of the detectors behind focusing telescopes.  The energy spectrum of Sco X-1 can be decomposed into several components: a thermal component possibly from the inner accretion disk or the neutron star surface, an iron emission line, and a dominant Comptonization component \citep{Barnard2003,Bradshaw2003,Dai2007,Church2012,Titarchuk2014}.  Based on X-ray dipping and other evidence for the corona size, an extended accretion disk corona model is proposed for X-ray Comptonization in LMXBs \citep{Church2004} as well as in Sco X-1 \citep{Barnard2003,Church2012}.  On the other hand, spectral analysis reveals that the seed photons for Comptonization may have a temperature exceeding the maximum temperature of the accretion disk. Thus, the seed photons likely originate from the neutron star surface, and consequently, the corona is located in the transition layer \citep{Dai2007,Titarchuk2014}. This scenario is favored by studies with frequency resolved spectroscopy \citep{Revnivtsev2006}.  It is also suggested that a corona in the transition layer can up-scatter seed photons from both the disk and neutron star surface \citep{Titarchuk2014}. 

X-ray polarimetry is sensitive to the geometry in radiation transfer, and may help distinguish different corona models \citep{Schnittman2010}.  The Bragg polarimeter on OSO-8 observed Sco X-1 in the 1970s \citep{Long1979}, and produced a non-detection around 2.6~keV but a 3$\sigma$ measurement around 5.2~keV with a polarization fraction (PF) of $0.0131 \pm 0.0040$ and a polarization angle (PA) of $57^\circ \pm 6^\circ$.  Launched in 2018, PolarLight is the second dedicated astrophysical X-ray polarimeter in the keV band \citep{Feng2019}. We thereby conducted polarization measurements of Sco X-1 with PolarLight after we finished observations of the Crab nebula \citep{Feng2020a,Long2021}. In this paper, we report a significant detection of polarization and discuss how the result can help constrain the location and geometry of the corona in Sco X-1.

\section{Observations and analysis}

\begin{figure}
\centering
\includegraphics[width=\columnwidth]{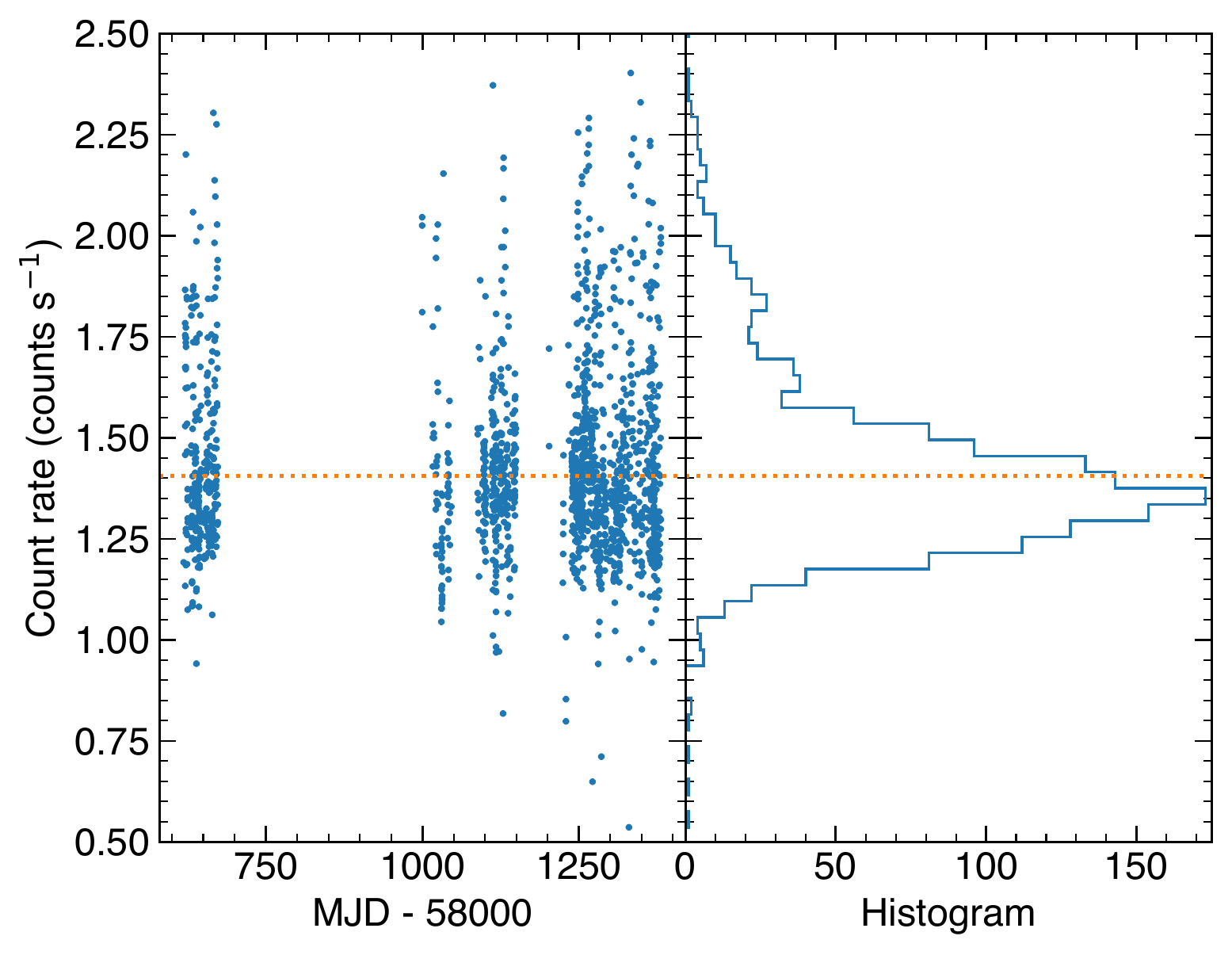}
\caption{Intensity of Sco X-1 in each exposure as a function of time measured with PolarLight in the energy range of 2--8 keV and a histogram of the intensity. The dashed line marks the level to separate the low- and high-intensity exposures, with approximately the same number of counts in each segment.
\label{fig:lc}}
\end{figure}

Sco X-1 has been observed with PolarLight in four time epochs, from 2019-May-14 to 2019-July-09,  2020-May-30 to 2020-July-16, 2020-August-27 to 2020-October-29, and 2021-January-10 to 2021-June-15, with a total exposure of 884~ks. The first two observing windows were scheduled during the Sun avoidance for observations of the Crab nebula.  In August 2020, the monitoring program for the Crab nebula ended and Sco X-1 became the primary target of PolarLight. In late 2020, the schedule was interrupted in response to a target of opportunity.  After mid-June 2021, the instrument turned to observe another target of opportunity.  A continuous exposure with PolarLight typically lasts 10 minutes.  The exposure-by-exposure lightcurve is shown in Figure~\ref{fig:lc} with a histogram of the source intensity (count rate). 

The data reduction is the same as that employed in the analysis for the Crab observations \citep{Long2021}.  A valid event contains an image of energy deposit, i.e., the track image. The background events are screened using an energy dependent algorithm \citep{Zhu2021}.  Some background events are produced by secondary electrons with an energy close to the X-rays of interest, and are thus not removable. Events with at least 58 pixels and located in the central $\pm7$~mm region of the detector are selected for analysis. 

The energy of X-rays, or precisely the relation between the energy and analog-to-digital numbers, may vary with time and is calibrated by comparing the measured and simulated source spectra \citep{Li2021}. For this purpose, one needs to assume an incident source spectrum. However, Sco X-1 displays spectral variability that may affect the accuracy of the calibration.  \citet{Titarchuk2014} performed a comprehensive spectral study of the source with RXTE observations at different spectral branches. Fortunately, the variation in spectral shape in our energy band is not dramatic. We plot all these spectral models and find the ``typical'' one that has a median spectral slope in 2--8 keV. This one catches the source in the normal branch. Then, we take the 5th and 95th percentile spectra ordered by the spectral slope as two extremes. The X-ray energies calibrated against the typical spectrum and the two extremes differ by a factor less than 10\% in 2--8 keV. This is less than the energy resolution (23\% around 2 keV and 16\% around 8 keV, full width at half maximum) of the detector and can be ignored. 

\begin{figure*}
\centering
\includegraphics[width=0.43\textwidth]{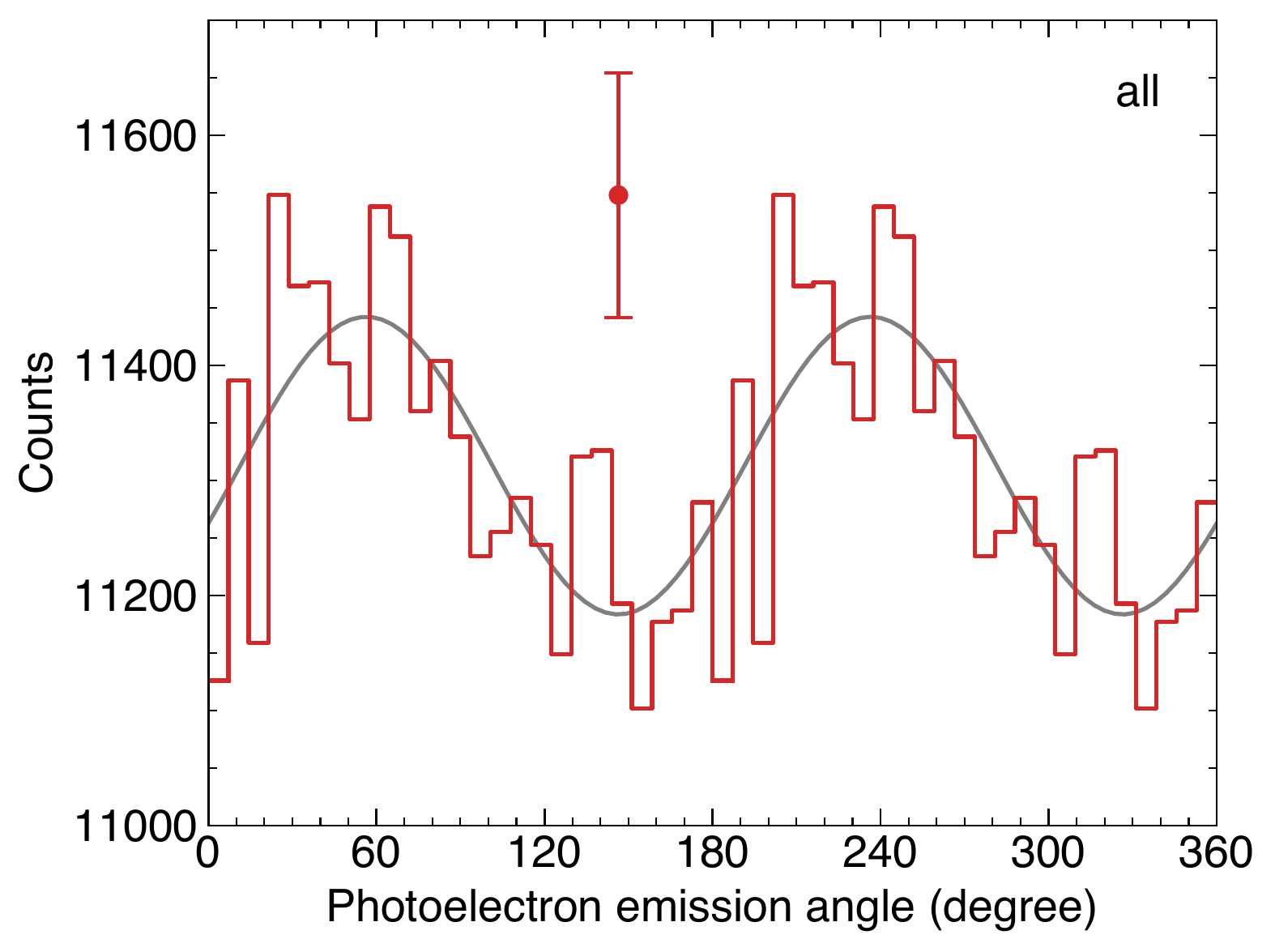}
\includegraphics[width=0.43\textwidth]{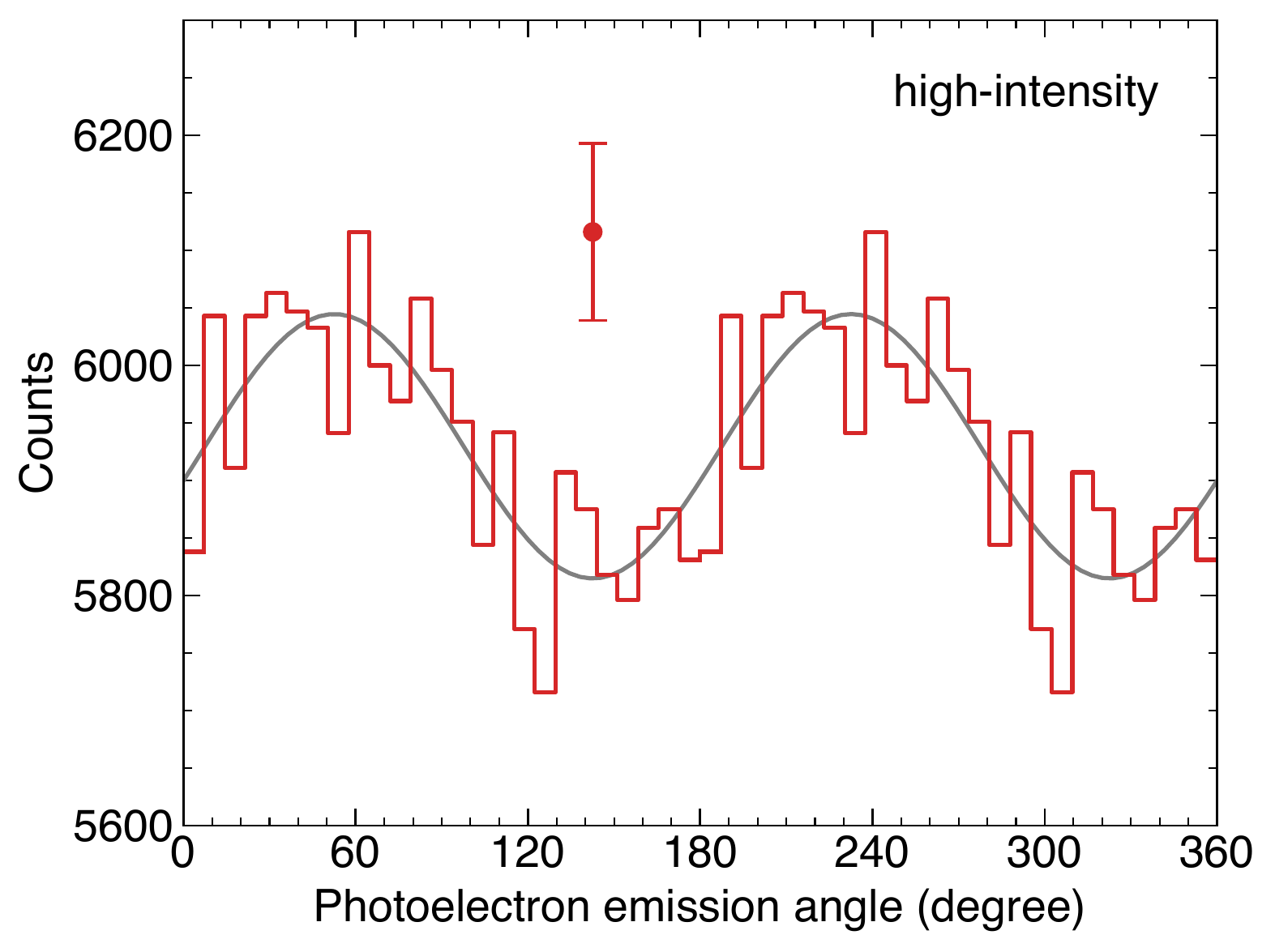} \\
\includegraphics[width=0.4\textwidth]{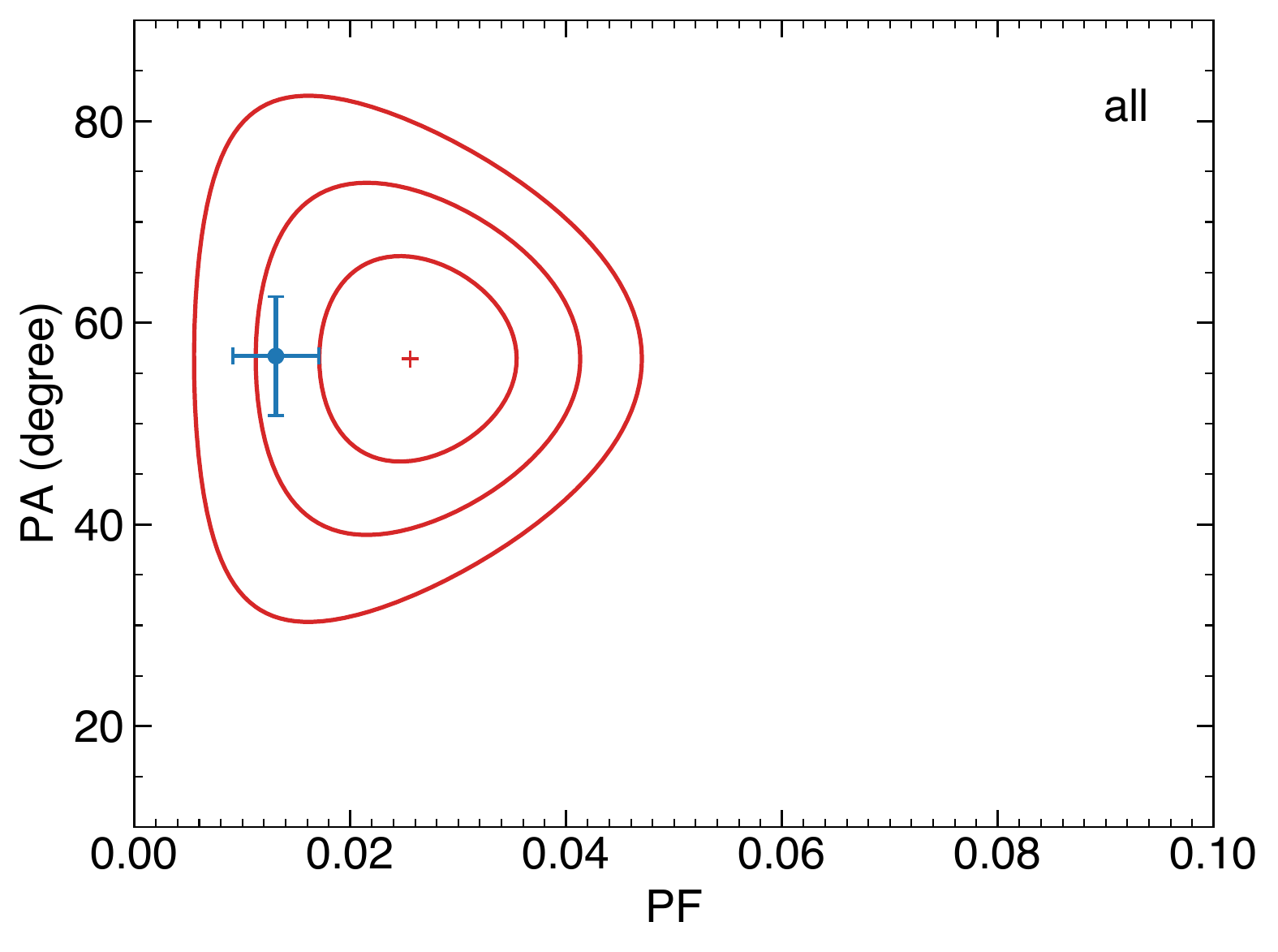}
\hspace{3mm}
\includegraphics[width=0.4\textwidth]{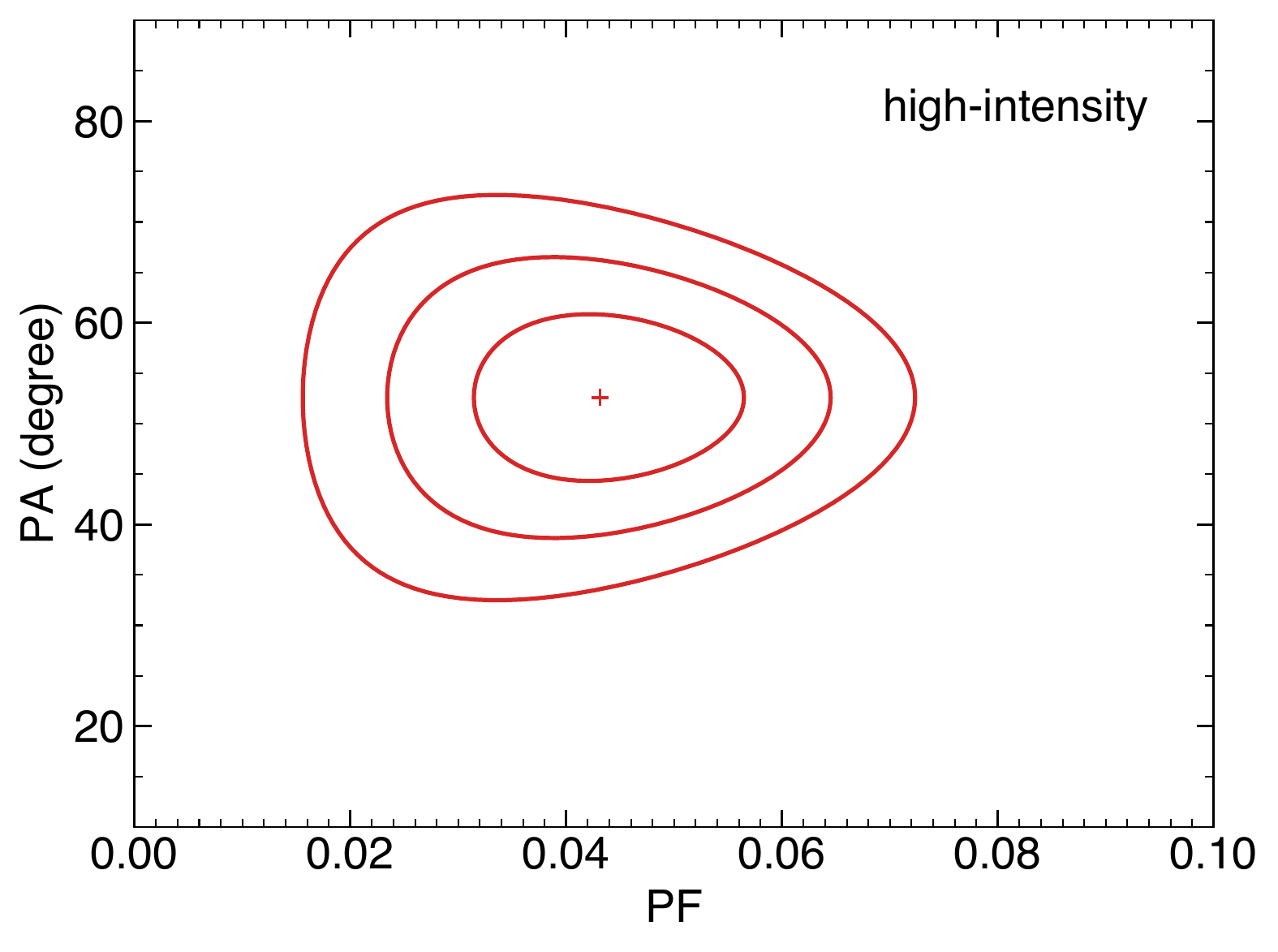}
\caption{Polarimetric modulation curves (red histograms) and corresponding PA vs.\ PF contours of Sco X-1 in the 4--8 keV band, in all and the high-intensity time intervals, respectively. {\bf Top}: the modulation curves are for visual inspection only,  with typical 1$\sigma$ error bars,  and the model curves (grey lines) are derived from the Stokes/Bayesian analysis. {\bf Bottom}: the red crosses indicate the point estimates and the contours encircle the 1$\sigma$, 2$\sigma$, and 3$\sigma$ credible intervals of the Bayesian posterior distribution.  The blue point with error bars marks the measurement obtained with OSO-8 around 5.2 keV. 
\label{fig:mod}}
\end{figure*}

The emission angle of photoelectrons is inferred using the impact point method \citep{Bellazzini2003}.  The average modulation factors in different energy bands (see Table~\ref{tab:pol}) are calculated based on the laboratory calibrations \citep{Feng2019} weighted by the measured source spectrum. The background fraction during on-source observations is estimated to be on the order of 1\%, also listed in Table~\ref{tab:pol}.  Identical to previous studies \citep{Feng2020a,Long2021}, we calculate the polarization based on the Stokes parameters \citep{Kislat2015,Mikhalev2018}, and infer the intrinsic PF and PA using a Bayesian approach \citep{Maier2014,Mikhalev2018} that is immune from the bias in polarization analysis.  The marginalized posterior distribution is used for parameter estimates.  The credible interval is calculated as the region with the highest posterior density.

\begin{deluxetable}{cclllll}
\centering
\tabletypesize{\footnotesize}
\tablecaption{X-ray polarization measurements of Sco X-1 with PolarLight.}
\label{tab:pol}
\tablehead{
\colhead{Energy} & \colhead{Intensity} & \colhead{$N_{\rm ph}$} & \colhead{$f_{\rm b}$} & \colhead{MDP} & \colhead{PF} & \colhead{PA} \\
\colhead{(keV)} & & & & & & \colhead{(\arcdeg)} \\
\colhead{(1)} & \colhead{(2)} & \colhead{(3)}  & \colhead{(4)} & \colhead{(5)} & \colhead{(6)}  & \colhead{(7)}}
\startdata
3--4 & all & 353558 & 0.007 & 0.022 & $0.000_{-0.000}^{+0.011}$ & $40.0 \pm 29.3$ \\
4--8 & all & 282822 & 0.017 & 0.018 & $0.026_{-0.006}^{+0.006}$ & $56.4 \pm 6.7$ \\
\noalign{\smallskip}\hline\noalign{\smallskip}
3--4 & low & 176211 & 0.008 & 0.031 & $0.008_{-0.008}^{+0.009}$ & $5.2 \pm 26.6$ \\
4--8 & low & 134579 & 0.019 & 0.026 & $0.000_{-0.000}^{+0.011}$ & $77.8 \pm 36.1$ \\
\noalign{\smallskip}\hline\noalign{\smallskip}
3--4 & high & 177347 & 0.006 & 0.031 & $0.018_{-0.012}^{+0.010}$ & $62.2 \pm 18.1$ \\
4--8 & high & 148243 & 0.015 & 0.025 & $0.043_{-0.008}^{+0.008}$ & $52.6 \pm 5.4$ \\
\enddata
\tablecomments{
Column~(1): Energy range.
Column~(2): Intensity range. 
Column~(3): Total number of events used for polarization analysis.
Column~(4): Fraction of background events estimated in this energy and intensity interval. 
Column~(5): The minimum detectable polarization (MDP) at 99\% confidence level, $= 4.29 / [\mu (1 - f_{\rm b})\sqrt{N_{\rm ph}} ]$, where $\mu$ is the modulation factor, or the fractional modulation amplitude in response to fully polarized X-rays, about 0.33 in 3--4 keV and 0.46 in 4--8 keV.
Column~(6): Polarization fraction.
Column~(7): Polarization angle in degree.
Errors are quoted as 68\% credible intervals.
}
\end{deluxetable} 

The polarization measurement is performed in two energy bands, 3--4 and 4-8 keV, respectively.  The analysis in 3--4 keV yields a PF consistent with zero, while the PF measured in 4--8 keV has a significance of 4$\sigma$. Due to the small effective area and relatively narrow band of PolarLight, it is not possible to calculate the color-color or hardness-intensity diagram for branch identification on the timescales suited for that purpose. The data therefore cannot be categorized based on position on the Z-track. Instead, we opt to divide the data into two segments (low/high intensity) based solely on the intensity in the exposure, in a way that produces approximately the same number of counts in both segments.  This ensures that the two segments have almost the same sensitivity in polarization measurement.  As Sco X-1 exhibits similar intensities in the horizontal and normal branches, but a higher intensity in the flaring branch \citep{Titarchuk2014}, the two datasets can still be expected to correlate to the position in the Z-track. 

In the 4--8 keV energy range, the low-intensity half shows a PF consistent with zero, while a 5$\sigma$ detection is obtained with the high-intensity data, with ${\rm PF} = 0.043 \pm 0.008$ and ${\rm PA} = 52\fdg6 \pm 5\fdg4$.  The significance that we quote is purely statistical, and is verified to be consistent with the sensitivity (MDP) of the observation given the number of photons and modulation factor \citep{Weisskopf2010,Strohmayer2013}.  The polarization measurements in different energy bands and intensity intervals are listed in Table~\ref{tab:pol}.  The modulation curves and PA vs.\ PF contours are shown in Figure~\ref{fig:mod}. Our measurement in the 4--8 keV band in all intensity intervals is consistent with that obtained with OSO-8 around 5.2~keV.  

\subsection{Arguments against background or systematic effects}

\begin{figure}
\centering
\includegraphics[width=\columnwidth]{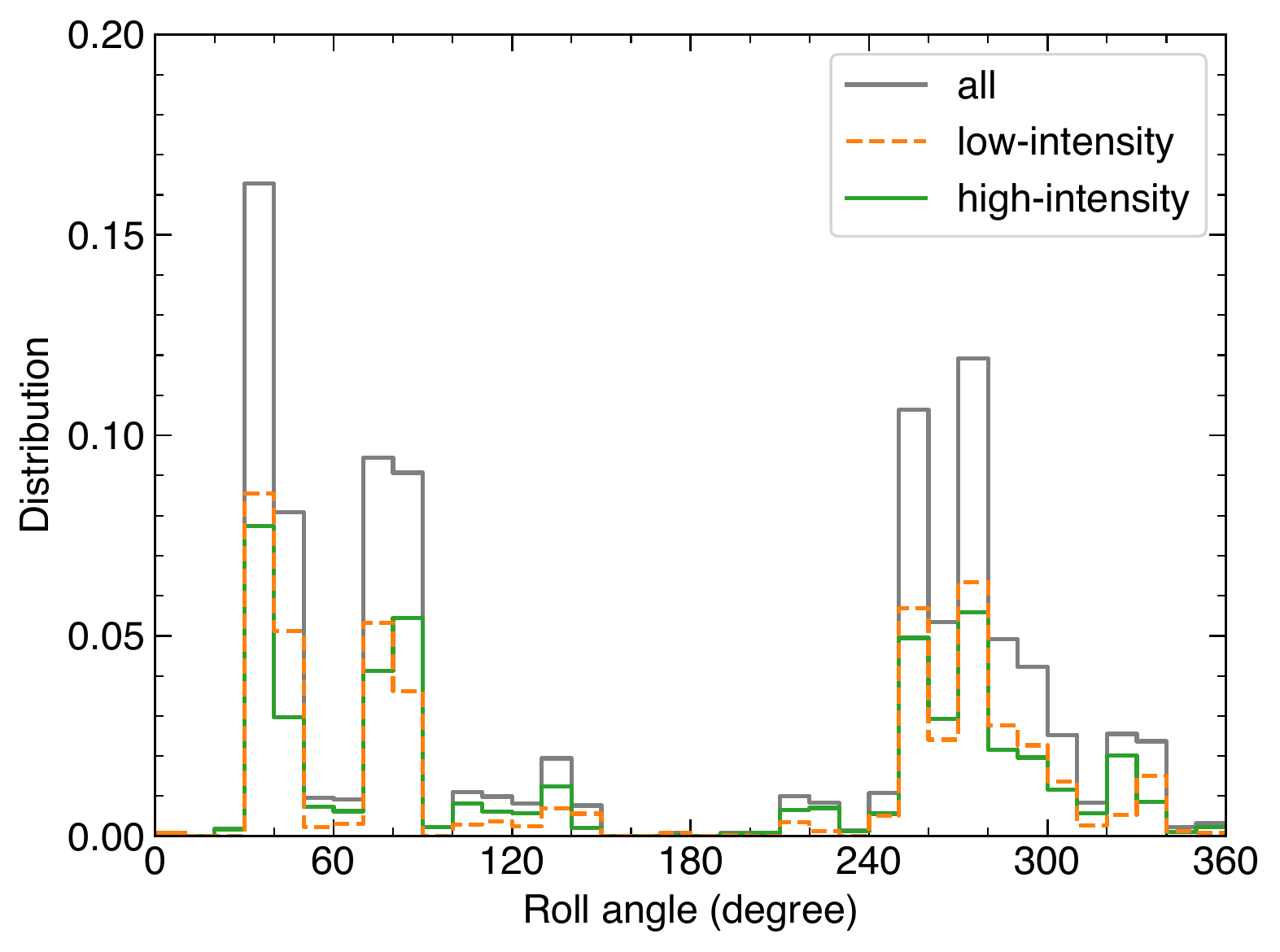}
\caption{Distribution of the PolarLight roll angle during observations of Sco X-1, and those when the source displays low and high intensities.
\label{fig:roll}}
\end{figure}

We present evidence that the polarization measurements are not affected by the background or instrument systematics. The background count rate is too low for us to constrain any spurious modulation in it.  The background fraction is estimated to be about 1.5\% in 4--8 keV during high-intensity intervals.  Even if it is 100\% polarized, it cannot produce a PF as high as 0.043 as observed. Furthermore, if the detection is due to background effects, it should be more significant at low-intensity intervals rather than in high-intensity intervals. To conclude, a significant background effect on the detection can be ruled out. 

Unpolarized X-rays may result in a residual modulation in a pattern similar to that caused by polarization. Such an instrument systematics at 5.9~keV is lower than 1\% averaged over the detector plane for this type of detectors \citep{Li2015}.  For PolarLight, the response to an unpolarized beam, 5.9~keV from an $^{55}$Fe source, has been found to be low, with a 90\% upper limit of 0.016 on PF. Any instrument rotation, such as from the roll pattern shown in Figure~\ref{fig:roll}, would further suppress such a residual modulation. We find that, by folding with the roll pattern in either the low- or high-intensity intervals, the amplitude of modulation is lowered by a factor of 2.  Considering that the low- and high-intensity datasets have the same energy range, comparable sensitivity for polarization measurements, and similar roll angle distributions, it is unlikely for a systematic effect from the unpolarized response to appear significantly in one dataset but not in the other.  Also, the instrument systematics, if any, is only significant in the energy band below 4 keV \citep{Baldini2021}; a non-detection in the low energy band while a significant detection in the high energy band does not reconcile with the behavior of such an effect. 

Furthermore, the independent measurements with PolarLight and OSO-8 produce a consistent PA within errors, $52\fdg6 \pm 5\fdg4$ vs.\ $57\arcdeg \pm 6\arcdeg$. This further strengthens the reliability of the results, as the PA measurement is a pure geometric effect, almost not affected by calibration uncertainties.  

\section{Discussion}

\begin{figure}[t]
\centering
\includegraphics[width=\columnwidth]{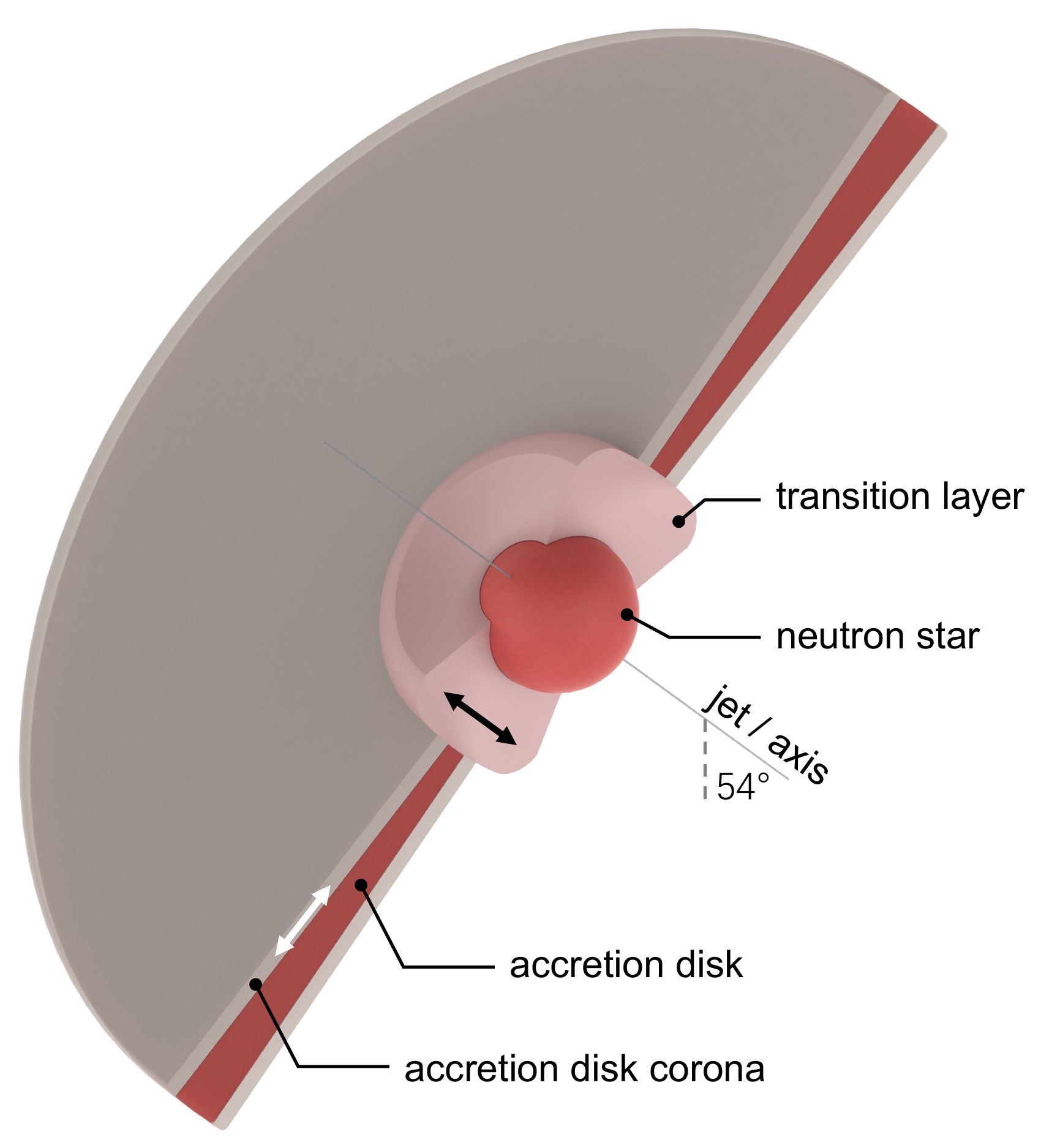}
\caption{Schematic drawing of the accretion flow and expected polarization angles of the two models. The radio jet has an orientation of 54$^\circ$ on the sky plane and is likely the symmetric axis of the accretion flow. Given an optically thin corona, if the Comptonization occurs in the accretion disk corona, the PA is expected to be aligned with the disk plane, as indicated by the white arrow; otherwise, if the Comptonization occurs in the transition layer, the PA is expected to be in line with the system axis, as indicated by the black arrow. 
\label{fig:drawing}}
\end{figure}

The measured PA is in line with the orientation ($54^\circ$ from north to east) of the radio jet on the sky plane \citep{Fomalont2001}. It is reasonable to assume that the jet is perpendicular to the inner accretion disk.  Therefore, the PA is perpendicular to the disk, i.e., parallel with the symmetric axis of the system.  In the energy range of a few keV, the flux fraction from the Comptonization component increases with energy \citep{Dai2007}. A non-detection at low energies and a significant detection toward high energies suggests that the thermal emission is intrinsically of low polarization and the signal is a result of scattering in the corona.  Spectral modeling uncovers that the corona of Sco X-1 is optically thin at the highest luminosities \citep{Dai2007}, with an optical depth below 1 in some of the flaring branch and as low as 0.1 on the top of the flaring branch.  In this case, the average number of scatters is less than unity, and the PA can be inferred from the geometry as follows. If the corona is flat and extended above the disk, with seed photons from the disk beneath, a PA perpendicular to the system axis is expected. Otherwise, if the corona resides in the transition layer, then the PA is expected to be parallel with the system axis or the radio jet.  A physical picture of the accretion system is illustrated in Figure~\ref{fig:drawing}.  

The signature of X-ray polarization with a wedge like corona above the accretion disk (sandwich geometry) around a stellar mass black hole has been studied with simulation \citep{Schnittman2010}. This is similar to the geometry of the accretion disk corona discussed above, and may be used for an informative comparison despite a neutron star in Sco X-1. The polarization is found to be parallel with the disk plane if the photons are scattered once, but perpendicular to it with multiple scatters \citep{Schnittman2010}.  Considering the low optical depth and a seed photon temperature of 2--3~keV  \citep[corresponding to a thermal peak at 5--8~keV;][]{Dai2007}, the majority of the scattered photons in 4--8 keV should be scattered only once in the corona \citep{Rybicki1986}, leading to the same conclusion as discussed above. Simulations assuming an active galactic nucleus \citep{Beheshtipour2017} may not be applicable to our case because the seed photon temperature is much lower. If there are multiple scatters, although the PA fits the measurement, their simulations indicate that the PF decreases with increasing luminosity \citep{Schnittman2010}, opposite to what we have found. 

A spherical corona with seed photons from a truncated disk is also investigated \citep{Schnittman2010}. This is slightly different from the geometry of a transition layer corona, where the spherical caps are removed and the thermal emission from the neutron star surface could be another important seed (see Figure~\ref{fig:drawing}).  A sphere without caps breaks the symmetry and can help increase the observed PF.  The average PA after scattering in a spherical corona is expected to be perpendicular to the disk, and the perpendicular component mainly originates from the portion of corona at low latitudes and relatively large radii, i.e.,  in the region occupied by the transition layer corona \citep[see Figure~12 in][]{Schnittman2010}.  Therefore, the simulation results agree with our first principle analysis. To conclude, the polarization measurements favor the scenario that an optically thin corona is located in the transition layer when the source undergoes high accretion rates. 

Nevertheless, to better understand the physics and geometry, dedicated simulations are needed. This work implies that, especially for the study of systems with low strength of magnetic fields, it is important to have time and energy resolved X-ray polarimetry, as well as broadband spectroscopy to determine the emission state, to break the degeneracy in the model. These can be fulfilled with future missions \citep{Weisskopf2016,Zhang2019}.

\begin{acknowledgments}
HF acknowledges funding support from the National Natural Science Foundation of China (grants Nos.\ 11633003, 12025301, and 11821303), the CAS Strategic Priority Program on Space Science (grant No.\ XDA15020501-02), and the National Key R\&D Project (grants No.\ 2018YFA0404502).  
\end{acknowledgments}



\end{document}